\documentclass[aps,prl,twocolumn,superscriptaddress,groupedaddress]{revtex4-1}
\usepackage[compat=1.0.0]{tikz-feynman}

\usepackage{amssymb}
\usepackage{multirow}
\usepackage{bm}
\usepackage{amsmath}
\usepackage{graphicx}
\usepackage{epstopdf}
\usepackage{subfigure}
\usepackage{natbib}
\usepackage{epsfig}
\usepackage{amsfonts}
\usepackage{mathrsfs}
\usepackage[toc,page,title,titletoc,header]{appendix}
\usepackage[colorlinks,linkcolor=blue,citecolor=blue,anchorcolor=blue]{hyperref}

\usepackage{dsfont,amsthm,amsbsy}

\def\tr{{\rm tr}}

\usepackage{fancyhdr}

\usetikzlibrary{quotes,angles}
\newcommand{\obtuseangle}{\kern.08em
\begin{tikzpicture}
    \draw coordinate (a) at (0.14,0);
    \draw coordinate (b) at (0,0);
    \draw coordinate (c) at (-.12,0.18);
    \draw (a) -- (b) -- (c) pic [draw=black]{} ;
\end{tikzpicture}%
\kern.08em%
}

\newcommand{\plaquette}{\kern.08em
\begin{tikzpicture}[baseline={([yshift=-1ex]current bounding box.center)}]
    \draw coordinate (a) at (0,0);
    \draw coordinate (b) at (0.1,0);
    \draw coordinate (d) at (0.06,0.09);
    \draw coordinate (c) at (0.16,0.09);
    \draw (a) -- (b) -- (c) -- (d) -- (a) pic [draw=black]{} ;
\end{tikzpicture}%
\kern.08em%
}

\newcommand{\plaquettedimerone}{\kern.08em
\begin{tikzpicture}[baseline={([yshift=-.5ex]current bounding box.center)}]
    \draw coordinate (a) at (0,0);
    \draw coordinate (b) at (0.35,0);
    \draw coordinate (d) at (0.2,0.33);
    \draw coordinate (c) at (0.55,0.33);
    \draw[color=black!60] (a) circle (0.04);
    \draw[color=black!60] (b) circle (0.04);
    \draw[color=black!60] (c) circle (0.04);
    \draw[color=black!60] (d) circle (0.04);
    \draw[black,very thin] (a) -- (b);
     \draw[black,very thin] (c) -- (d);
    \draw[black, ultra thick] (b)--(c);
    \draw[black, ultra thick] (d)--(a);
\end{tikzpicture}%
\kern.08em%
}

\newcommand{\plaquettedimertwo}{\kern.08em
\begin{tikzpicture}[baseline={([yshift=-.5ex]current bounding box.center)}]
    \draw coordinate (a) at (0,0);
    \draw coordinate (b) at (0.35,0);
    \draw coordinate (d) at (0.2,0.33);
    \draw coordinate (c) at (0.55,0.33);
    \draw[color=black!60] (a) circle (0.04);
    \draw[color=black!60] (b) circle (0.04);
    \draw[color=black!60] (c) circle (0.04);
    \draw[color=black!60] (d) circle (0.04);
    \draw[black,ultra thick] (a) -- (b);
     \draw[black,ultra thick] (c) -- (d);
    \draw[black, very thin] (b)--(c);
    \draw[black, very thin] (d)--(a);
\end{tikzpicture}%
\kern.08em%
}

\newcommand{\hexagon}{\kern.08em
\begin{tikzpicture}[baseline={([yshift=-.5ex]current bounding box.center)}]
    \draw coordinate (a) at (0.08,0);
    \draw coordinate (b) at (0.04,0.07);
    \draw coordinate (c) at (-0.04,0.07);
    \draw coordinate (d) at (-0.08,0.);
    \draw coordinate (e) at (-0.04,-0.07);
    \draw coordinate (f) at (0.04,-0.07);
    \draw (a) -- (b) -- (c) -- (d) -- (e) -- (f) -- (a) pic [draw=black]{} ;
\end{tikzpicture}%
\kern.08em%
}

\newcommand{\trimer}{\kern.08em
\begin{tikzpicture}[baseline={([yshift=-.5ex]current bounding box.center)}]
    \draw coordinate (a) at (-0.25,0.2);
    \draw coordinate (b) at (0,0);
    \draw coordinate (c) at (0.25,0.2);
    \draw[color=black!60] (a) circle (0.04);
    \draw[color=black!60] (b) circle (0.04);
    \draw[color=black!60] (c) circle (0.04);
    \draw[black,double,thin] (a) -- (b);
    \draw[black,double,thin] (b) -- (c);
\end{tikzpicture}%
\kern.08em%
}

\newcommand{\monomerdimer}{\kern.08em
\begin{tikzpicture}[baseline={([yshift=-.5ex]current bounding box.center)}]
    \draw coordinate (a) at (0,0);
    \draw coordinate (b) at (0.25,-0.2);
    \draw coordinate (c) at (0.5,0);
    \draw[color=black!60] (a) circle (0.04);
    \draw[color=black!60] (b) circle (0.04);
    \draw[color=black!60] (c) circle (0.04);
    \draw[black,very thin] (a) -- (b);
    \draw[black, ultra thick] (b)--(c);
\end{tikzpicture}%
\kern.08em%
}

\newcommand{\dimermonomer}{\kern.08em
\begin{tikzpicture}[baseline={([yshift=-.5ex]current bounding box.center)}]
    \draw coordinate (a) at (0,0);
    \draw coordinate (b) at (0.25,-0.2);
    \draw coordinate (c) at (0.5,0);
    \draw[color=black!60] (a) circle (0.04);
    \draw[color=black!60] (b) circle (0.04);
    \draw[color=black!60] (c) circle (0.04);
    \draw[black,very thin] (c) -- (b);
    \draw[black, ultra thick] (a)--(b);
\end{tikzpicture}%
\kern.08em%
}

\newcommand{\threesites}{\kern.08em
\begin{tikzpicture}[baseline={([yshift=-1ex]current bounding box.center)}]
    \draw coordinate (a) at (0,0);
    \draw coordinate (b) at (0.1,-0.08);
    \draw coordinate (c) at (0.2,0);
    \draw (a) -- (b) -- (c) pic [draw=black]{} ;
\end{tikzpicture}%
\kern.08em%
}

\newcommand{\twosites}{\kern.08em
\begin{tikzpicture}[baseline={([yshift=-1ex]current bounding box.center)}]
    \draw coordinate (a) at (0,0);
    \draw coordinate (b) at (0.12,0);
    \draw (a) -- (b) pic [draw=black]{} ;
\end{tikzpicture}%
\kern.08em%
}

\newcommand{\monomermonomer}{\kern.08em
\begin{tikzpicture}[baseline={([yshift=-.5ex]current bounding box.center)}]
    \draw coordinate (a) at (0,0);
    \draw coordinate (b) at (0.3,0);
    \draw[color=black!60] (a) circle (0.04);
    \draw[color=black!60] (b) circle (0.04);
    \draw[black,very thin] (a) -- (b);
\end{tikzpicture}%
\kern.08em%
}

\newcommand{\Ueph}{U_{\text{e-ph}}}
\newcommand{\lij}{\langle ij\rangle}

\usepackage{ulem}

\begin{document}
\title{Resonating valence bond states %and other surprises 
in an electron-phonon system}
\author{Zhaoyu~Han}
\affiliation{Department of Physics, Stanford University, Stanford, California 94305, USA}
\author{Steven~A.~Kivelson}
\affiliation{Department of Physics, Stanford University, Stanford, California 94305, USA}

\begin{abstract}
We study a simple electron-phonon model on square and triangular versions of the Lieb-lattice using an asymptotically exact strong coupling analysis. At zero temperature and electron density $n = 1$ (one electron per unit cell), for various ranges of parameters in the model, we exploit a mapping to the quantum dimer model to establish the existence of a spin-liquid phase with $\mathbb{Z}_2$ topological order (on the triangular lattice) and a multi-critical line corresponding to a quantum critical spin liquid (on the square lattice). In the remaining part of the phase diagram, we find a host of charge-density-wave phases (valence-bond solids), a conventional s-wave superconducting phase, and with the addition of a small Hubbard $U$ to tip the balance, a phonon-induced d-wave superconducting phase. Under a special condition, we find a hidden pseudo-spin $SU(2)$ symmetry that implies an exact constraint on the superconducting order parameters.
\end{abstract}
\maketitle

The electron-phonon interaction plays an essential role in the physics of quantum materials, e.g. for Bardeen-Cooper-Schrieffer superconductivity (SC) and typical charge or bond-density-wave ordering~\cite{PhysRev.108.1175,PhysRevLett.42.1698,RevModPhys.60.781}. In the past few years, it has become increasingly clear that electron-phonon interactions can also induce a variety of more exotic behaviors and novel quantum phases~\cite{PhysRevLett.121.247001,PhysRevLett.125.167001,huang2021pair,PhysRevB.105.L100509,PhysRevB.100.245105,PhysRevB.106.235112,PhysRevLett.127.247203,PhysRevB.105.085151,PhysRevB.106.L081115,PhysRevB.106.245136,PhysRevB.51.3840,PhysRevB.51.505, PhysRevLett.77.723, PhysRevB.54.14971, PhysRevB.53.431, PhysRevB.75.014503, PhysRevB.83.064521,  PhysRevLett.121.257001,PhysRevResearch.2.023006}, including those that are typically associated with strong repulsive interactions, e.g. anti-ferromagnetism (AF)~\cite{PhysRevLett.127.247203,PhysRevB.105.085151,PhysRevB.106.L081115,PhysRevB.106.245136} and d-wave SC~\cite{PhysRevB.51.3840,PhysRevB.51.505, PhysRevLett.77.723, PhysRevB.54.14971, PhysRevB.53.431, PhysRevB.75.014503, PhysRevB.83.064521,  PhysRevLett.121.257001,PhysRevResearch.2.023006}.

\begin{figure}[b]
    \centering
    \includegraphics[width=0.95 \linewidth]{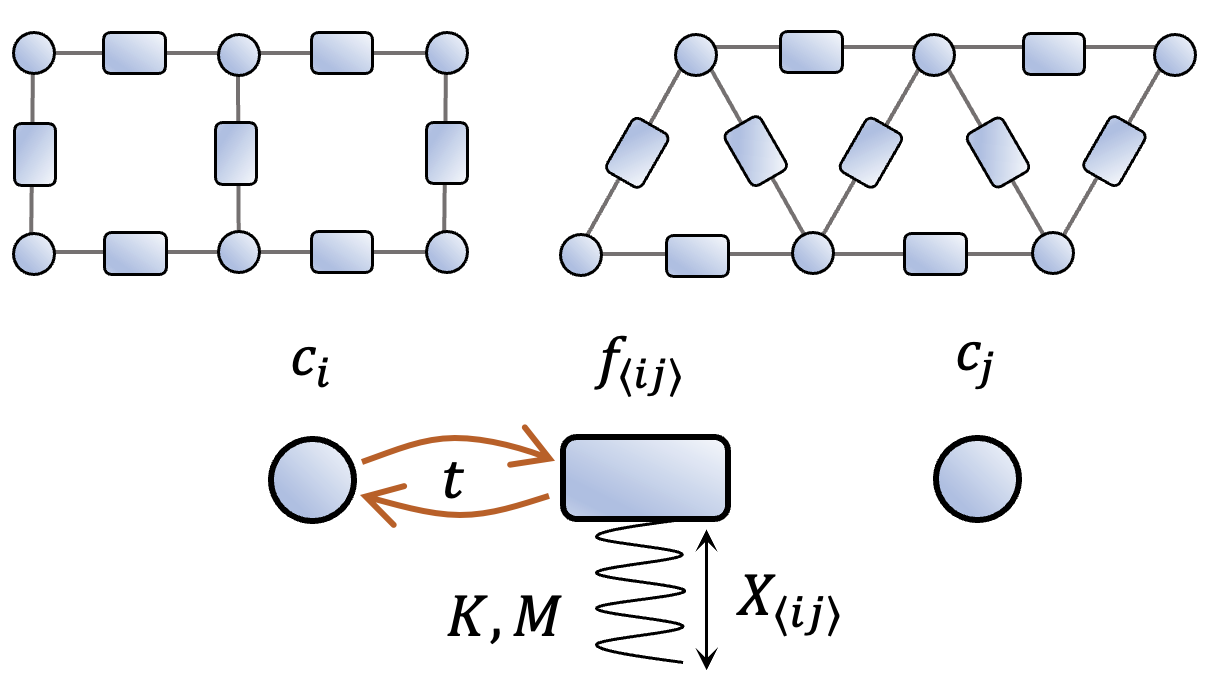}
    \caption{An illustration of the Lieb lattices studied in this paper, and the ``Holstein-Lieb'' model in Eq.~\ref{originalHamiltonian}. }
    \label{ModelIllustration}
\end{figure}

\begin{figure}[t]
    \centering
    \includegraphics[width = \linewidth]{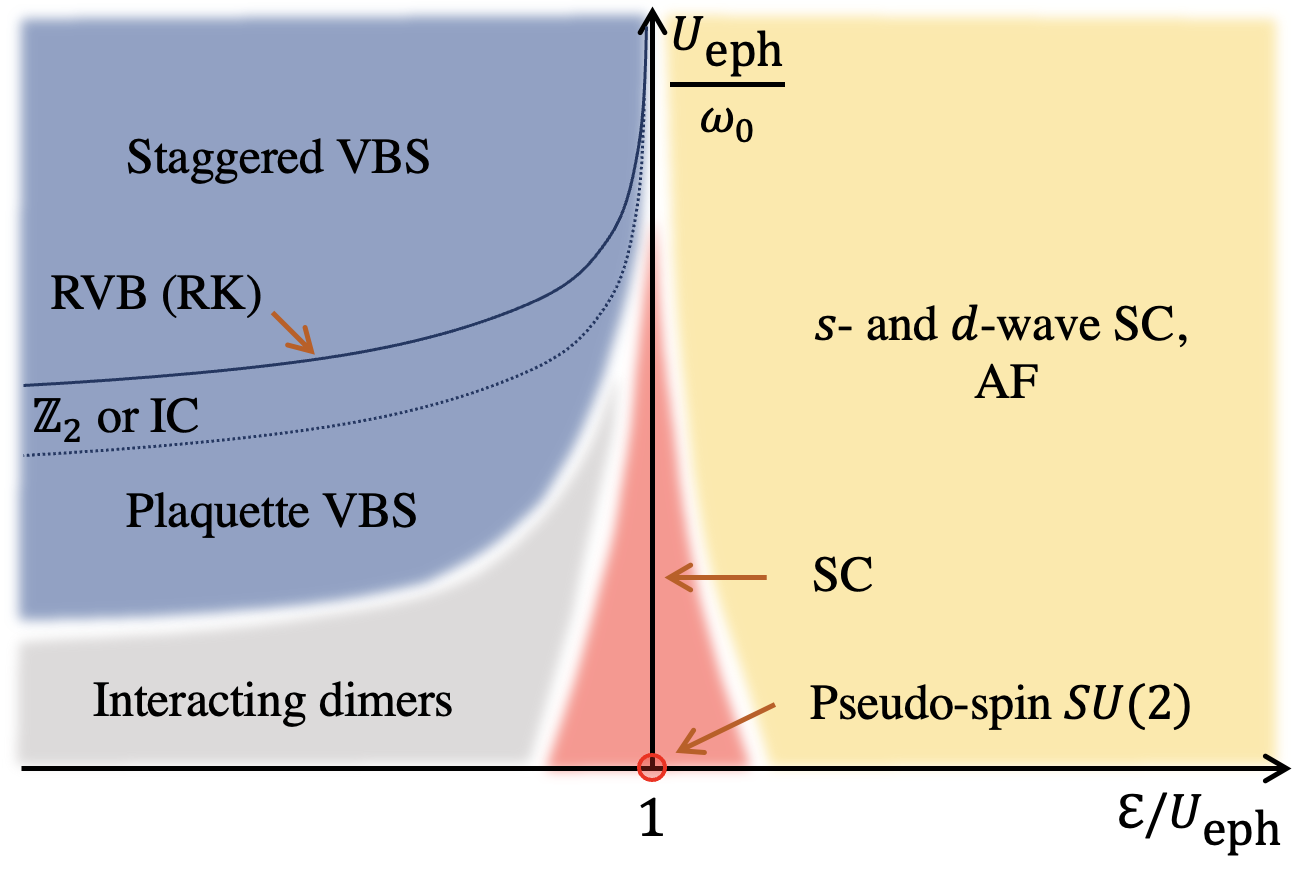}
    \caption{Schematic %quantum 
    $T=0$ phase diagram for $n 
    =1$ in the small $|t|$ limit as a function of the     dimensionless ratios of parameters in the model,  
    Eq. \ref{originalHamiltonian}; the vertical axis quantifies the degree of retardation and the horizontal axis the strength of the electron-phonon coupling.
    The blue region on the left top is described by the   quantum dimer model in Eq.~\ref{RK}. The yellow region on the right is described by the weakly interacting theory in Eq.~\ref{cHamiltonian}. The red region in the middle is described by Eq.~\ref{BandF} and is confined to a narrow window. The grey region on the left bottom is described by the effective bosonic Hamiltonian, Eq.~\ref{BHamiltonian}, where controlled analysis of the ground-state phases is missing. The RK line,  which corresponds to the exactly solvable point $V_2=\tau_2$ in Eq.~\ref{RK}, occurs with moderately large retardation, $\Ueph/\omega_0 = \mathcal{O}\left( \ln\frac{\mathcal{E}_{12}}{t}\right)$, as long as  $\mathcal{E}/\Ueph$ is not close to $1$. The $\mathbb{Z}_2$ symbol represents a $\mathbb{Z}_2$ spin liquid phase on the triangular lattice, and IC stands for possible incommensurate crystalline phases on the square lattice. }
    \label{PhaseDiagram}
\end{figure}

In this letter, we study a simple electron-phonon model, the ``Holstein-Lieb'' model (illustrated in Fig.~\ref{ModelIllustration}), for which it is possible to obtain well-controlled results concerning the ground-state phase diagram (summarized in Fig.~\ref{PhaseDiagram}) through the use of an asymptotic strong-coupling expansion. Certain of the phases  are interesting but not surprising -- for instance, phonon-stabilized bipolarons can order (localize) to form a variety of valence bond solid (VBS) phases, or when the strongly coupled sites lie above the Fermi energy, they act as ``negative $U$ centers'' that mediate SC pairing~\cite{geballe2016paired}.  More unexpectedly, there is a range of parameters in which the problem maps onto a quantum dimer model~\cite{PhysRevLett.61.2376,moessner2011quantum} introduced by Rokshar and Kivelson (RK),  and thus exhibits a variety of exotic ``resonating valence-bond'' (RVB) phases known to arise there, including (on the triangular lattice) a $\mathbb{Z}_2$ topologically ordered phase and (on the square lattice) a multicritical point which acts as the mother-state for an infinite hierarchy of incommensurate phases. With this concrete example, we hope to suggest new avenues for the search for materials supporting ``spin liquid'' phases~\cite{savary2016quantum} in systems with relatively strong electron-phonon couplings. To date, this effort has been almost entirely focused on studies of frustrated anti-ferromagnets.

%We also find a region of strong -- albeit subdominant -- d-wave pairing tendency, such that a primarily phonon-driven d-wave superconducting state can be stabilized in the presence of a small  electron-electron repulsion represented by a Hubbard $U$. 

We consider a Hamiltonian in which electrons can occupy orbitals on the vertices of the lattice, $j$, or on the bond centers between pairs of nearest-neighbor sites, $\langle ij\rangle$:  
\begin{align}\label{originalHamiltonian}
    \hat{H} = & - t \sum_{\langle ij \rangle \sigma }  \left[\hat{f}^\dagger_{\langle ij \rangle\sigma} \left(\hat{c}_{i\sigma}+\hat{c}_{j\sigma}\right)+ \text{h.c.}\right]  \nonumber \\
    & + \sum_{\langle ij \rangle } \left(\mathcal{E} + \alpha \hat{X}_{\langle ij\rangle}\right) \hat{n}_{\langle ij\rangle} +\hat{H}_\text{ph} 
\end{align}
where $\hat{c}_{i\sigma}$ ($\hat{f}_{\langle ij\rangle \sigma}$) annihilates a spin-$\sigma$ electron on the orbital at site $i$ (bond $\langle ij\rangle $), and $\hat{n}_i$ or $\hat{n}_{\lij}$ are the electron numbers on the corresponding orbitals. $\hat{X}_{\langle ij\rangle}$ are the coordinate operators of optical phonons on bonds described by the Hamiltonian:
\begin{align}
    \hat{H}_\text{ph} = \sum_{\lij}\left[\frac{K \hat{X}^2_{\langle ij \rangle}}{2} + \frac{ \hat{P}^2_{\langle ij \rangle}}{2M}\right]\ .
\end{align}
By rescaling variables it is easy to see that there are precisely four independent energy scales in the problem:  the band-structure scale $t$, the charge transfer gap $\mathcal{E}$, the phonon frequency, $\omega_0\equiv \sqrt{K/M}$, and the phonon-induced electronic attraction, $\Ueph \equiv \alpha^2/K$,  which is the relevant measure of the electron-phonon coupling strength. We will perform a controllable strong-coupling analysis assuming $|t|$ to be a small energy scale (the precise meaning of this assumption will be made clear below), and obtain effective theories for the active degrees of freedom in different parameter regimes.  At the end of the paper, we will discuss  the robustness of the results in the presence of additional couplings that  are likely to be present in candidate materials.

 %Indeed, at the end of the paper we will discuss several aspects of generalizing the model to include direct Hubbard interactions, 
% \begin{align}
%     \hat { H}_\text{Hubbard} \equiv   U_c \sum_{i} \hat{n}_{i\uparrow} \hat{n}_{i\downarrow} +U_f \sum_{\lij } \hat{n}_{\lij \uparrow} \hat{n}_{\lij \downarrow}
% \end{align}
% which can be analyzed in the same controlled manner.

{\bf Methods. } The effects of strong electron-phonon coupling can best be addressed following a unitary transformation~\cite{lang1963kinetic}
\begin{align}
\hat{U}\equiv \exp \left[\mathrm{i}\frac{\alpha}{K}\sum_{\langle ij\rangle} \hat{P}_{\langle ij\rangle} \hat{n}_{\langle ij\rangle} \right] \ , 
\end{align} 
that transforms the Hamiltonian into:
\begin{align}\label{transformedHamiltonian}
    \hat{U}^\dagger \hat{H} \hat{U}
    = & - t \sum_{\langle ij \rangle \sigma }  \left[ \hat{D}_{\langle ij \rangle} \hat{f}^\dagger_{\langle ij \rangle\sigma} \left(\hat{c}_{i\sigma}+\hat{c}_{j\sigma}\right)+ \text{h.c.}\right]  \nonumber \\
    & \ \ + \sum_{\langle ij \rangle } \left[\mathcal{E} \hat{n}_{\langle ij\rangle} - \frac{\Ueph}{2} \hat{n}_{\langle ij\rangle}^2\right] + \hat{H}_\text{ph}
\end{align}
% \begin{align}\label{transformedHamiltonian}
%      \hat{H} = & - t \sum_{\langle ij \rangle \sigma }  \left[ \hat{f}^\dagger_{\langle ij \rangle\sigma} \left(\hat{c}_{i\sigma}+\hat{c}_{j\sigma}\right)+ \text{h.c.}\right]  \nonumber \\
%     & \ \ + \sum_{\langle ij \rangle } \left[\mathcal{E} \hat{n}_{\langle ij\rangle} + \frac{U_f}{2} \hat{n}_{\langle ij\rangle}^2\right] +\sum_i \frac{U_c}{2} \hat{n}_{i}^2
% \end{align}
The resulting theory contains a unitary operator $\hat{D}_{\langle ij \rangle}\equiv\mathrm{e}^{-\mathrm{i} \hat{P}_{\langle ij\rangle} \alpha / K }$ that displaces the phonon coordinate by $\alpha/K$. We will perform a perturbative analysis treating the second line in Eq.~\ref{transformedHamiltonian} as the unperturbed Hamiltonian, $\hat{H}_0$, and the first line as perturbations, $\hat{H}'$. Under most circumstances, $\hat{H}_0$ has an extensive ground-state degeneracy, so we use degenerate perturbation theory to derive an effective model acting in the degenerate subspace. {  We note that all the virtual processes, including those with phonon excitations, are included in this analysis. } We keep terms in powers of $t$ to the lowest order needed to resolve the degeneracy, and the validity of each model will be analyzed in each scenario. Since $\hat{H}$ can be defined on any lattice in any dimension, so can the resulting effective theories. Here, to be explicit, we will confine ourselves to the square and triangular lattices in two dimensions. 

The first step in our analysis is to identify the degenerate ground-state manifold of $\hat{H}_0$, which we call $\mathcal{H}_0$; we will restrict our attention to the range of electron densities per unit cell, $0<n\leq 2$. %~\footnote{The full range of possible $n$ runs from 0 to $n=2+z$ where $z$ is the  number of nearest-neighbor bonds radiating from a site.}
For convenience, we define $\mathcal{E}_1 \equiv \mathcal{E} - \Ueph/2$ and $\mathcal{E}_2 \equiv 2\mathcal{E} - 2\Ueph$ to represent the energies of a singly or doubly occupied bond orbital, and $\mathcal{E}_{12} \equiv \mathcal{E}_1-\mathcal{E}_2 = 3\Ueph/2 - \mathcal{E}$ to represent their energy difference. Because $\mathcal{E}_2 - 2\mathcal{E}_1 =-\Ueph$ is always negative, singly occupied bond orbitals are always disfavored. Therefore, depending on the sign of $\mathcal{E}_2$, 
bond or site orbitals  are favored, 
so that all  
possible occupation configurations of bond dimers or site electrons form a basis of $\mathcal{H}_0$.  

%When $\mathcal{E}_2$ is positive ($\Ueph < \mathcal{E}$), electrons tend to stay on $c$ (site) orbitals. In this case, $\mathcal{H}_0$ consists of all possible electron occupation configurations on sites. When $\mathcal{E}_2$ is negative ($\Ueph > \mathcal{E}$), the electrons tend to stay on $f$ (bond) orbitals in pairs; it is convenient to associate a pair of electrons on bond $\langle ij\rangle$ with a dimer  occupying this bond.  In this language, a basis for $\mathcal{H}_0$ can be labeled by dimer occupation configurations on bonds. When $\mathcal{E}_2 \approx 0$, both orbitals are active, the analysis is more complicated as will be discussed later. 

For the case where bond dimers are active degrees of freedom ($\mathcal{E}_2\lesssim 0$), 
there are two sorts of terms that will be generated by the perturbative analysis:   There are diagonal terms (dimer potential energy) and off-diagonal terms (dimer kinetic energy). Since the phonon displacements are different for different dimer configurations, all off-diagonal terms must vanish in the limit of non-dynamical phonons, $\omega_0= 0$. Specifically, the amplitude of any process in which a dimer relocates onto or off of a bond  is accompanied by a Frank-Condon factor $F$~\cite{carlson2008concepts} defined as
\begin{align}
    F \equiv \langle 0| \hat{D}^2 | 0 \rangle = \mathrm{e}^{-X} , 
\end{align} 
where $|0\rangle$ is the ground state of the phonon Hamiltonian on the bond, and $X\equiv \Ueph/\omega_0$ is a dimensionless factor quantifying the degree of retardation, and the displacement operator $\hat{D}$ is squared since the occupancy of the orbital changes by two electrons. This factor becomes arbitrarily small in the limit of strong retardation. On the other hand, the potential terms always only receive $\mathcal{O}(1)$ factors from the  virtual phonon fluctuations.

Below we derive the effective theories and  obtain expressions as functions of the bare energy scales for the coupling constants that arise in low-order perturbation theory in $t$. The effective Hamiltonians are given in Eqs.~\ref{BHamiltonian},~  \ref{cHamiltonian}~\&~\ref{BandF}. The asymptotic expressions and the limiting behaviors in the small and large $\omega_0$ limits of the effective couplings are given in Table.~\ref{effectivecoefficients}. Their explicit expressions and derivations are deferred to Supplemental Materials~\footnote{See Supplemental Materials and Refs.~\cite{PhysRevLett.125.167001, PhysRevLett.63.2144, PhysRevLett.65.120, agterberg2020physics} therein for explicit derivations and expressions for the effective coefficients, the detailed discussion on the pseudo-spin symmetry, and the mean-field analysis of the weakly interacting model.}.

\begin{table}[t] 
    \centering
\begin{tabular}{| c | c | c |}
\hline
 & anti-adiabatic  & adiabatic  \\
 & $\omega_0 \rightarrow \infty $, $X\rightarrow 0 $ & $\omega_0 \rightarrow 0$ , $X \rightarrow \infty$ \\
  \hline
$t_\text{eff} \sim \frac{t^2}{\mathcal{E}_1}$ & $\frac{t^2}{\mathcal{E}_1}$ & $\frac{t^2}{\mathcal{E}}$  \\
\hline
$\tau_0 \sim \frac{2t^2 F }{\mathcal{E}_1} $ & $\frac{2t^2}{\mathcal{E}_1}$ & $\frac{ \sqrt{\pi X} t^2 }{\mathcal{E}_1} \mathrm{e}^{-X}\rightarrow 0$ \\
  \hline
 $\tau_1 \sim \frac{t^4 F^2}{\mathcal{E}_{12}^3}$ & $\frac{4t^4 (2\Ueph-\mathcal{E})}{\mathcal{E}_{12}^2|\mathcal{E}_{2}|\Ueph}$  & $\frac{4 \sqrt{2\pi X } t^4}{\mathcal{E}_{2}^2 \Ueph}\mathrm{e}^{-2X}\rightarrow 0$\\ 
 \hline
 $V_1 \sim \frac{t^4}{\mathcal{E}_{12}^3}$  & $\frac{4t^4(5\Ueph/2-\mathcal{E})}{\mathcal{E}_{12}^3\Ueph} $ &  $\frac{2t^4(4\Ueph-\mathcal{E})}{(2\Ueph- \mathcal{E})^3\Ueph} $\\
\hline
$J_\text{eff} \sim \frac{t^4}{\mathcal{E}_1^3}$ &$\frac{4t^4 \Ueph}{\mathcal{E}_1^3 \mathcal{E}_2}$ &  $\frac{t^4 \Ueph^3 }{2 \mathcal{E}_1^2 \mathcal{E}^4}$ \\
 \hline
\end{tabular}
     \caption{Asymptotic expressions and limiting behaviors of the coefficients in the effective theories in Eqs.~\ref{BHamiltonian}, \ref{cHamiltonian}~\&~\ref{BandF}.  Each limiting behavior  is evaluated in the region of  validity of the corresponding 
     effective Hamiltonian.
     }
    \label{effectivecoefficients}
\end{table}

{\bf $\Ueph > \mathcal{E}$: dimer models. } In this case, the energy necessary for breaking a dimer is $\mathcal{E}_{12}$; thus the expansion series in $t$ is controllable as long as $t\ll \mathcal{E}_{12}$. For a dimer on bond $\lij$, we define the annihilation operator as $\hat{b}_{\lij} \equiv \hat{f}_{\lij\uparrow} \hat{f}_{\lij\downarrow}$ and the dimer occupation number operator $\hat{n}^b_{\lij} \equiv \hat{b}^\dagger_{\lij}\hat{b}_{\lij} = 0 , 1$. To the fourth order in $t$, we obtain the following model for the dimers:
\begin{align}\label{BHamiltonian}
\hat{H}_{b} =  \sum_{\langle i j k \rangle}\left[ -\tau_1 \left(\hat{b}_{\langle ij\rangle }^\dagger \hat{b}_{\langle jk \rangle} + \text{h.c.} \right) + V_1 \hat{n}^b_{\langle ij \rangle} \hat{n}^b_{\langle jk \rangle}\right]
\end{align}
where the summation is over all pairs of nearest-neighbor bonds with a single vertex in common, and it is implicit that we have omitted terms of order $t^6$ and higher, to which we shall return shortly. Therefore, we obtain a hard-core boson model on the bond lattice with  repulsive interactions.

As shown in Table.~\ref{effectivecoefficients}, when $X\lesssim 1$ it follows that $\tau_1 \sim V_1$, so this is an interacting problem with no small parameter to give theoretical control. We label this region
``Interacting dimers'' in Fig.~\ref{PhaseDiagram}. For $n\ll 1$, the ground state is presumably an s-wave superfluid, independent of the details of the lattice structure and the interactions.  As $n$ approaches 1, the balance between the interactions and the kinetic terms becomes more subtle.
Similar models have been studied on several lattices~\cite{PhysRevLett.84.1599, PhysRevLett.88.167208,PhysRevB.77.014524,PhysRevLett.95.127207,PhysRevLett.102.017203,PhysRevB.75.174301}, where various superfluid, charge and supersolid orders were found.  We expect analogous phases  to arise in the present model.

\begin{figure}[b]
    \centering
    \includegraphics[width=0.8 \linewidth]{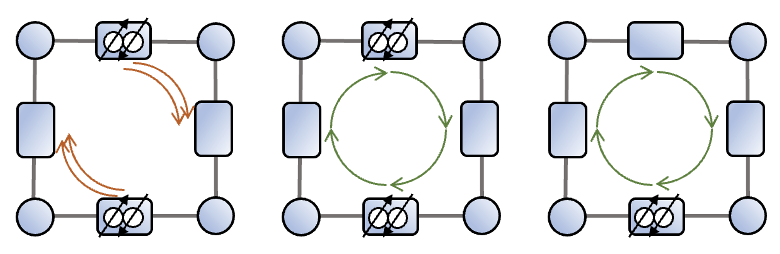}
    \caption{ 
    Illustration of the virtual processes (arrows indicate the direction of electron hops) on a four-sided plaquette contributing to the terms in the RK effective theory in Eq.~\ref{RK}. The first class contributes to $\tau_2$ while the latter two contribute to $V_2$. \label{RKillustration} }
\end{figure}

However, when $X$ is large, such that $F\ll 1$, the $V_1$ term is dominant over $\tau_1$. Those ground states of $V_1$ within $\mathcal{H}_0$ form an emergent low-energy (still extensively degenerate) subspace, $\mathcal{H}_1$, in which, as in the RK quantum dimer models, no more than one dimer can touch a given site.  Within $\mathcal{H}_1$, we further perform perturbative analysis on square and triangular lattices and obtain the RK model as the effective model:
\begin{align}\label{RK}
    \hat{H}_{\text{RK}} =&   V_2 \sum_{\plaquette}\left[ \left|\plaquettedimerone 
    \right\rangle \left\langle \plaquettedimerone \right| + \left|\plaquettedimertwo
    \right\rangle \left\langle \plaquettedimertwo \right| \right] \nonumber\\
    & -\tau_2 \sum_{\plaquette}\left[ \left|\plaquettedimerone 
    \right\rangle \left\langle \plaquettedimertwo \right| + \left|\plaquettedimerone
    \right\rangle \left\langle \plaquettedimertwo \right| \right]
\end{align}
where the ket (bra) represents a pair of annihilation (creation) operators on the thickened bonds, and the summation is over all possible four-sided plaquettes. To leading order in $\tau_1/V_1$, it is easy to see that $\tau_2 =\frac{4 \tau_1^2}{V_1} \sim \frac{t^4}{\mathcal{E}_{12}^3} F^4$.  In terms of the same expansion, one would conclude that $V_2 = \mathcal{O}\left(\tau_1^5/V_1^4\right)$ is always small compared to $\tau_2$. However, since we are simultaneously assuming both $t$ and $F$ are small, we need to consider terms (not shown in Eq.~\ref{BHamiltonian}) that are higher order in $t$, but which are not suppressed by $F$ (potential terms). {  (See Fig.~\ref{RKillustration} for illustrations of the virtual processes that contribute to the effective theory.)  } Since it involves such high-order processes, it is not worth writing out explicitly the results. What is essential is that the leading term only contributes to $V_2$ and it is eighth order in $t$ and positive, i.e. $V_2  \sim \frac{t^8}{\mathcal{E}_{12}^7}$~\footnote{As an aside, we note that a $\pi$ external magnetic flux per plaquette will turn this repulsion into attraction.}. From the perturbative perspective, there are two types of leading virtual process, both of which contribute positively to $V_2$: If two dimers occupy two parallel sides of a four-sided plaquette, the lowest order virtual process that non-trivially connects them is a ring exchange of electrons, which results in a fermion minus sign that turns what would have been an energy gain into a cost. On the other hand, if only one dimer occupies a side of a plaquette, there is an allowed  virtual process in which a single electron travels around the plaquette - this energy-gaining process is blocked when there are two dimers on the plaquette. Both processes are eighth order in $t$ on the lattices we are considering.
[We have performed exact diagonalization studies on small clusters in the limit $\omega_0=0$ as a check of these conclusions.]

Formally, we can consider $\hat{H}_{\text{RK}}$ to be the effective Hamiltonian in an asymptotic limit where $F\sim |t|/\mathcal{E}_{12}\sim \delta \ll 1$, such that the  couplings $\tau_2$ and $V_2$ are both of order $\delta^8$, with relative magnitudes that can be tuned in a wide range - for instance by varying $\omega_0$. All omitted terms are higher order in powers of $\delta$.

The zero-temperature phase diagrams of the quantum dimer models Eq.~\ref{RK} on the square and triangular lattices have been solved, with results we briefly summarize here.  For both cases,  the line along which $\tau_2=V_2$ is special (corresponding to the RK point~\cite{PhysRevLett.61.2376,moessner2011quantum}) and is labeled ``RK'' in Fig.~\ref{PhaseDiagram}. Here, an exact ground state is an equal amplitude superposition of all dimer configurations in a given topological sector, i.e. a short-ranged RVB state~\cite{savary2016quantum}. The ground states in different topological sectors are exactly degenerate, which leads to topological degeneracy on compact manifolds. Through exact evaluations of the dimer correlations~\cite{PhysRevB.66.214513,PhysRev.132.1411,fradkin2013field}, it is known that this point is gapless on the square lattice and gapped on the triangular lattice. 

Further numerical and analytical investigations have fleshed out the full phase diagram of this model. On the square lattice,  for the model defined in Eq. \ref{RK}, the RK point is a critical point separating two different VBS states: staggered (for $V_2>\tau_2>0$) and plaquette (for $0<V_2<\tau_2$)~\cite{PhysRevLett.100.037201,PhysRevB.103.094421}. 
More generally, it is an unstable multi-critical point described by the quantum Lifshitz model~\cite{PhysRevB.69.224415,fradkin2013field}, near which a wide class of perturbations can induce incommensurate crystalline phases% around the point
 through a mechanism known as Cantor deconfinement~\cite{PhysRevB.69.224415,PhysRevB.75.094406}. We label the region that may host such additional phases ``IC'' in Fig.~\ref{PhaseDiagram}.~\footnote{
  We note that these relevant terms generically exist in the higher-order terms that are neglected in this study. It is also worth noting that the inclusion of a small admixture of further (second neighbor) range dimers can stabilize a $\mathbb{Z}_2$ spin liquid phase proximate to the RK point, similar to that seen in the triangular lattice~\cite{PhysRevLett.99.247203}. %Therefore, slightly away from the asymptotic limit, the RK point is likely to broaden into a regime with multiple additional quantum phases. %Interestingly, it is also possible that a power-law critical phase may also be stabilized near a range of parameters around the RK point at finite temperature~\cite{PhysRevB.64.144416,dabholkar2022re}.}
 }
 On the triangular lattice, the RK point lies on the boundary of a phase %regime 
 that exhibits $\mathbb{Z}_2$ topological order in a range of $\nu_c \tau_2 <V_2<\tau_2$ with $\nu_c \lesssim 0.8$ (marked %with 
$\mathbb{Z}_2$ in Fig.~\ref{PhaseDiagram}); two different VBS occur for other ranges of parameters: staggered (for $V_2>\tau_2>0$) and $\sqrt{12}\times \sqrt{12}$ (for $0<V_2<\nu_c \tau_2$)~\cite{PhysRevLett.86.1881,PhysRevB.71.224109}, which we also refer to as ``plaquette VBS'' in the schematic phase diagram Fig.~\ref{PhaseDiagram}.

The $\mathbb{Z}_2$ spin liquid is known to have several types of excitations: spinons, holons, and visions. In the current case, while visions have relatively low creation energy $\sim \tau_2$, the creation of spinons or holons  carries a large energy cost $\sim \mathcal{E}_2$ or $\sim\Ueph$ in order to break a dimer. Therefore, when lightly doping holes into the system near the RK point such that $|n-1| = x \ll 1$, we will likely have dimer vacancies as charge carriers leading to condensation with SC $T_c$ determined by the coherence scale, $T_c \sim x \tau_1 \sim x \frac{t^6}{\mathcal{E}_{12}^5}$~\footnote{We note that there can be rich physics for dilute doped holes~\cite{PhysRevB.107.L140401,PhysRevB.102.214437, PhysRevB.107.L041103} even from an ordered state, which should be analyzed more carefully in future studies.}.
For the square lattice RK model, exactly at the RK point, there are also gapless ``resonon'' excitations  with momenta near $(\pi,\pi)$ and a quadratic dispersion. Since the motion of the dimers is tied to that of the phonons, the emergence of such excitations should be observable in measurements of the phonon spectrum, e.g. through neutron scattering.

{\bf $\Ueph < \mathcal{E}$: weakly interacting model. } In this case, the effective model is expressed in terms of site electrons. To fourth order in $t$, the effective Hamiltonian is
\begin{align}\label{cHamiltonian}
    \hat{H}_{c} = &- t_\text{eff} \sum_{\lij \sigma}\left(\hat{c}^\dagger_{i\sigma} \hat{c}_{j\sigma} + \text{h.c.} \right) - 2 J_\text{eff} \sum_{\lij }\hat{n}_{[i+j]\uparrow} \hat{n}_{[i+j]\downarrow}
\end{align}
where $\hat{n}_{[i+j]\sigma}
\equiv \left(\hat{c}^\dagger_{i\sigma}+ \hat{c}^\dagger_{j\sigma}\right)\left(\hat{c}_{i\sigma}+ \hat{c}_{j\sigma}\right)/2$ is the number of electrons in a bonding orbital between sites $i$ and $j$. Since any virtual movement of electrons necessarily costs $\mathcal{E}_1$ in the intermediate state, the expansion is valid as long as $|t|\ll \mathcal{E}_1$. %[Note that, this condition does not require $|t|\ll \Ueph$.] 
As can be seen from Table \ref{effectivecoefficients}, $t_\text{eff}$ is second order in $t$ and $J_\text{eff}$ is fourth order.  Thus,  we should consider this theory in its weak coupling limit, $J_\text{eff}\ll t_\text{eff}$. %This interaction term on each bond can be decomposed into a sum of an AF coupling, a pair hopping term, and a density attraction.

With detailed discussion in Supplemental Materials~\cite{Note1}, we analyze the weak-coupling instabilities in the context of a Hartree-Fock mean-field analysis that is reasonable in this limit.  We find that, on square and triangular lattices, $s$-wave SC is always the dominant instability for $n<2$. However, when $n\approx 1$ on the square lattice, there is also a $d$-wave pairing state that is only moderately subdominant to the dominant $s$-wave channel. This competition between the $s$- and $d$-wave paired states can be tuned by the additional weak Hubbard repulsion on site orbitals, $U_c$; when $U_c \gtrsim 2.2 J_\text{eff}$, the $d$-wave paired state has the lower variational energy. Exactly at $n=1$, there is also an AF instability, which is also subdominant to the $s$-wave SC instability, but which is favored over all superconducting states when $U_c > 2 J_\text{eff}$.

{\bf $\Ueph \approx \mathcal{E}$: monomer-dimer model. } In the narrow region  $|\mathcal{E}_2| \lesssim \tau_0$, where $\tau_0$ (again given in Table \ref{effectivecoefficients}) is the matrix element for converting a pair of site electrons to a pair of bond electrons, both $c$ and $f$ orbitals are active. The effective Hamiltonian in this case is
\begin{align}\label{BandF}
    \hat{H}_{{bc}} = & \sum_{\lij } \left[t_\text{eff} \left(2\hat{n}^b_{\lij} -1\right)  2\hat{n}_{[i+j]} +\left(\mathcal{E}_2-4t_\text{eff} \right) \hat{n}^b_{\lij} \right]\nonumber \\
&+\tau_0\sum_{\lij}\left[\hat{b}^\dagger_{\lij}\left(\hat{c}_{i\uparrow}+ \hat{c}_{j\uparrow}\right)\left(\hat{c}_{i\downarrow}+ \hat{c}_{j\downarrow}\right)+\text{h.c.}\right] 
\end{align}
 In this regime, no controllable analysis can be performed.  A mean-field analysis, treating $b$ and $c$ as decoupled, suggests an $s$-wave SC phase - presumably one that connects to the corresponding phase in the $\Ueph<\mathcal{E}$ case. 
%In the adiabatic limit where $\tau_0\rightarrow 0$, the boson configuration, $\{{n}^b_{\lij}\}$, is non-dynamical and the $c$ electrons are non-interacting. For this problem, we have performed a classical Monte Carlo study to find the ground-state boson configurations on the square and triangular lattices. For both cases, we find that for all $n\leq 2$, the electrons always tend to fill the site orbitals.  [For higher densities, we find a phase separation between a  dimer-free and a fully packed dimer region, until all the bonds are occupied.] This can be intuitively understood since the first line of Eq.~\ref{BandF} contains a $c$-electron-dimer repulsion. 

It is interesting to note that, for arbitrary $t$, there is a hidden pseudo-spin $SU(2)$ symmetry in the original problem when $\mathcal{E}_2=0$ and $X\rightarrow 0$ (marked with the red circle in Fig.~\ref{PhaseDiagram}), which is a generalization of that of the Hubbard model on bipartite lattices~\cite{PhysRevLett.63.2144,PhysRevLett.65.120}. % [If we include $U_c$ and $U_f$, the first condition becomes $\mathcal{E}+U_f/2 = \Ueph+U_c/2$.] 
This symmetry implies, in the thermodynamic limit, for the ground state of the system at {\it any filling}:
\begin{align}
    \frac{1}{N_\text{site}^2} \left \langle \left| \sum_i \hat{c}_{i\uparrow} \hat{c}_{i\downarrow} - \sum_{\lij } \hat{f}_{\lij \uparrow} \hat{f}_{\lij \downarrow}\right|^2\right\rangle = 0
\end{align}
which is, as discussed in detail in Supplemental Materials~\cite{Note1}, difficult to satisfy in any SC state that does not have space-dependent oscillations in sign.

%There are three possible physical conclusions from this result: 1. The ground state does not have off-diagonal long-range order (ODLRO)~\cite{RevModPhys.34.694}. 2. It has uniform ODLRO, but $2\langle \hat{c}_{i\uparrow} \hat{c}_{i\downarrow}\rangle=  \sum_j \langle \hat{f}_{\lij \uparrow} \hat{f}_{\lij \downarrow}\rangle$ is exact for any site $i$, where the sum is over all the neighboring sites. 3. It has ODLRO, but the pairing order parameters oscillate among different unit cells and average to zero, i.e. it is a pair density wave state~\cite{agterberg2020physics}. 

{\bf Outlook. }  The derivations of the effective theories can be easily generalized to include {\it strong} (in comparison to $t$) or even infinite Hubbard repulsion, $U_c$, on the site orbitals. In that case, the model corresponding to $\Ueph<\mathcal{E}$ at $n\lesssim 1 $  is a $t$-$J$ model (with no double occupancy constraint on site orbitals).  The AF coupling $J$ in this model 
is {\it enhanced} by the effective attraction on the bond orbitals, and an extra nearest-neighbor density-density repulsion interaction is induced by phonon virtual fluctuations. On the other hand, the dimer models for $\Ueph>\mathcal{E}$ are not qualitatively changed by the presence of a repulsive Hubbard interaction on site and bond orbitals, nor weak further-ranged hopping and electron-phonon coupling, as long as $\mathcal{H}_0$ is unaffected. 

In considering the search for ``spin liquid'' phases  in real materials featuring significant electron-phonon couplings, we summarize the key ingredients that we think are crucial for the mechanism revealed in this work: 1. atomic-scale structures with electronically active atoms on both vertices and the bridging sites between them (it is encouraging to note that a large class of real materials have this feature~\cite{FAN2022,yang2021ag,li2022lieb}); 2. strong coupling to phonon modes on bonds that allow the formation of bipolarons localized on bonds; 3. a moderately large degree of retardation that suppresses the quantum hopping relative to interactions and thus leads to  constraints on the low-energy Hilbert space. Furthermore, we would like to point out that these ideas (especially 1 and 3) can be adopted in the design of quantum simulation experiments as a novel way of realizing geometrical blockade analogous to the concept in Rydberg systems, which was crucial to a realization of ``spin liquid'' state in a recent experiment~\cite{PhysRevX.11.031005, doi:10.1126/science.abi8794}. In that context, the phonon degrees of freedom could be replaced by various other bosonic modes.

{\bf
Acknowledgment.} We thank Oskar Vafek, Ruben Verresen, Hong Yao, Kyung-Su Kim, and John Sous for helpful discussions. SAK was supported, in part, by NSF grant No. DMR-2000987 at Stanford.

\bibliographystyle{apsrev4-1} 
\bibliography{ref}

\onecolumngrid

\appendix

\section*{Resonating valence bond states in an electron-phonon system: Supplementary Materials }

\section{The effective coefficients}

\begin{table*}[b] 
    \centering
\begin{tabular}{| c | c | c |}
\hline
 & anti-adiabatic $\omega_0 \rightarrow \infty $ & adiabatic  $\omega_0 \rightarrow 0$ \\
  \hline
$t_\text{eff} = \frac{t^2}{\mathcal{E}_1} g_+(Y)$ & $\frac{t^2}{\mathcal{E} -\Ueph/2}$ & $\frac{t^2}{\mathcal{E}}$  \\
\hline
$\tau_0 = \frac{2t^2 F }{\mathcal{E}_1} g_-(Y) $ & $\frac{2t^2}{\mathcal{E} - \Ueph/2}$ & $\frac{\sqrt{\pi} t^2 }{\mathcal{E} - \Ueph/2} \sqrt{\frac{\Ueph}{\omega_0}}\mathrm{e}^{-\Ueph/\omega_0}\rightarrow 0$ \\
  \hline
 $\tau_1 = \frac{2t^4 F^2}{\mathcal{E}_{12}^2}\left[\frac{2}{|\mathcal{E}_2|} g^2_-(W ) + \frac{ 1}{\Ueph}  g_{3-}(W,X)\right]$ & $\frac{2t^4 (2\Ueph-\mathcal{E})}{(3\Ueph/2-\mathcal{E})^2(\Ueph-\mathcal{E})\Ueph}$  & $\frac{2t^4}{(\Ueph-\mathcal{E})^2}\sqrt{\frac{\pi}{2\Ueph \omega_0}}\mathrm{e}^{-2\Ueph/\omega_0}\rightarrow 0$\\ 
 \hline
 $V_1 = \frac{4t^4}{\mathcal{E}_{12}^2}\left[\frac{1}{\mathcal{E}_{12}} g_1(W) + \frac{1}{\Ueph} g_2(W,X) \right]$  & $\frac{4t^4(5\Ueph/2-\mathcal{E})}{(3\Ueph/2- \mathcal{E})^3\Ueph} $ &  $\frac{2t^4(4\Ueph-\mathcal{E})}{(2\Ueph- \mathcal{E})^3\Ueph} $\\
\hline
$J_\text{eff} = \frac{4t^4}{\mathcal{E}_1^2}\left[\frac{2}{\mathcal{E}_2} G_+(Y,Z) - \frac{1}{\mathcal{E}_{1}}g_1(Y)+\frac{1}{\mathcal{E}_1}G'_-(Y)\right]$ &$\frac{2t^4 \Ueph}{(\mathcal{E}-\Ueph/2)^3(\mathcal{E} - \Ueph)}$ &  $\frac{t^4 \Ueph^3 }{2 (\mathcal{E} -\Ueph/2)^2 \mathcal{E}^4}$ \\
 \hline
 \hline
 \multicolumn{1}{|c|}{ $g_\pm(\xi) \equiv  \xi \int_0^1 \frac{\mathrm{d}z}{z} ~  z^{\xi} \exp\left[ \pm \frac{X}{2}\left(z-1\right)\right] $ } &
  \multicolumn{2}{|c|}{$g_1(\xi) \equiv \xi^3     \int_0^1  \prod_{i=1}^3 \frac{\mathrm{d} z_i}{z_i} ~ (z_1z_2z_3)^{\xi} \exp\left[ \frac{X}{2}\left(z_1 z_2 +z_3 -2\right)\right] $ } \\
 \hline
 \multicolumn{3}{|c|}{ $g_2(\xi_1,\xi_2) \equiv \xi_1^2\xi_2 \int_0^1  \prod_{i=1}^3 \frac{\mathrm{d} z_i}{z_i}  ~ (z_1z_3)^{\xi_1} z_2^{\xi_2} \exp\left[\frac{X}{2}\left(z_1z_2z_3+z_2-2\right)\right] $}\\
  \hline
 \multicolumn{3}{|c|}{ $g_{3\pm}(\xi_1,\xi_2) \equiv \xi_1^2\xi_2 \int_0^1  \prod_{i=1}^3 \frac{\mathrm{d} z_i}{z_i}  ~ (z_1z_3)^{\xi_1 } z_2^{\xi_2} \exp\left[\pm \frac{X}{2}\left(z_1z_2+z_2z_3-2\right)\right] $}\\
 \hline 
 \multicolumn{3}{|c|}{$G_\pm(\xi_1,\xi_2) \equiv \xi_1^2\xi_2 \int_0^1  \prod_{i=1}^3 \frac{\mathrm{d} z_i}{z_i}  ~ (z_1z_3)^{\xi_1 } z_2^{\xi_2} \exp \frac{X}{2}\left[\mp \left(z_1+z_3\right) +z_2\left(1\pm z_1\right)\left(1\pm z_3\right)-2\right] $}\\
 \hline
\multicolumn{3}{|c|}{ $G_-'(\xi) \equiv \xi^3 \int_0^1  \prod_{i=1}^3 \frac{\mathrm{d} z_i}{z_i}  ~ (z_1z_3)^{\xi }  \exp\left[ \frac{X}{2}\left(z_1+z_3-2\right)\right] \left\{\exp\left[\frac{X}{2}z_2\left(1- z_1\right)\left(1-z_3\right)\right] - 1 \right\}$}\\
 \hline
\end{tabular}

     \caption{Expressions and  limiting behaviors of the coefficients in the effective theories in Eqs.~7, 9~\&~11 
     in terms of the dimensionless parameters $X \equiv \Ueph /\omega_0$, $Y \equiv \mathcal{E}_1/\omega_0$, $Z\equiv |\mathcal{E}_2|/\omega_0$, and $W \equiv \mathcal{E}_{12}/\omega_0$.  Each limiting behavior  is evaluated in the region of  validity of the corresponding 
     effective Hamiltonian.
     }
    \label{effectivecoefficients}
\end{table*}

For the second-order processes, it suffices to compute the following factor:
\begin{align}
    g_\pm (\xi) \equiv & \mathrm{e}^{(1\mp 1)X/2}\sum_{n} \frac{\langle 0 | D(\sqrt{X/2})|n\rangle \langle n| D(\mp \sqrt{X/2})|0\rangle }{1+n/\xi_1}\nonumber\\
    = & \xi \int_0^\infty \mathrm{d} t \exp \left[-\xi t \pm \frac{X}{2}\left(\mathrm{e}^{-t}-1\right)\right] 
\end{align}
where $D(\sqrt{X/2})$ is the displacement operator written in the coherent state basis. To obtain the second equality we have used the Feynman's trick $1/u = \int_0^\infty \mathrm{d} t \mathrm{e}^{-u t}$. Similar derivations can be found in the Supplemental Materials of Ref.~\cite{PhysRevLett.125.167001}.

Using these factors one can express
\begin{align}
    t_\text{eff} & = \frac{t^2}{\mathcal{E}_1} \sum_{n} \frac{\langle 0 | D(\sqrt{X/2})|n\rangle \langle n| D(\mp \sqrt{X/2})|0\rangle }{1+\omega_0 n/{\mathcal{E}_{1}}} = \frac{t^2}{\mathcal{E}_1} g_+(\mathcal{E}_{1}/\omega_0)
\end{align}
and similarly $\tau_0 = \frac{2t^2}{\mathcal{E}_1} g_-(\mathcal{E}_{1}/\omega_0) \mathrm{e}^{-X}$.

In anti-adiabatic limit $\omega\rightarrow \infty  $, $X, \xi\rightarrow 0 $, the $\mathrm{e}^{-t}$ term on the exponent of the integrand decays rapidly in the $t\sim 1/\xi$ scale, so we can neglect it and obtain $G\rightarrow \mathrm{e}^{-X/2} \sim 1$. In adiabatic limit $\omega\rightarrow 0 $, $X, \xi\rightarrow \infty $, the $\mathrm{e}^{-t}$ term on the exponent of the integrand decays slowly in the $t\sim 1/\xi$ scale, so we can expand it to $\mathrm{e}^{-t}\approx 1-t$ and obtain $G\rightarrow \frac{1}{1\pm X/(2\xi)}$. [For $g_-$, the above adiabatic result only applies when $X<2\xi$. When $X\approx 2\xi$, $g_- \rightarrow \sqrt{\pi X/4} $ in this limit. If $X>2\xi$, $g_- \rightarrow \xi \sqrt{4\pi / X }\mathrm{e}^{+X/4} $ will become an exponentially large factor that will give a correction to the Frank-Condon Factor $F(X) = \mathrm{e}^{-X}$. For the cases studied in this paper, only $X\approx 2\xi$ case is physically relevant. ]

For the fourth-order processes, it will be useful to evaluate the following quantity:
\begin{align}
    &I(\epsilon_1, \dots, \epsilon_4; a_1, \dots, a_4)\nonumber\\
    \equiv & \tr{\left[\sum_{\{n_i\}}  \bigotimes_{i=1}^4 \left(D(a_i)|n_i\rangle \mathrm{e}^{-\epsilon_i} \langle n_i|\right)\right]} \nonumber\\
      = & \int \frac{\prod_{i=1}^4 \mathrm{d} (\alpha_i,\alpha^\star_i) \mathrm{d} (\beta_i,\beta^\star_i)}{\pi^8} \exp\frac{-1}{2}\left[\sum_{i=1}^{4} |\alpha_i|^2+2|\beta_i|^2+|\alpha_i +a_i|^2 - 2\alpha^\star_i \beta_i \mathrm{e}^{-\epsilon_i} -a^\star_i \alpha^\star_i + a_i \alpha_i  - 2\beta^\star_i (\alpha_i + a_i)  \right]
\end{align}
where $|n\rangle$ is the $n$-th eigen state of the phonon Hamiltonian and $D(a) = \exp(a\hat{a}^\dagger - a^\star \hat{a})$ is the displacement operator. The second equality is obtained by inserting coherent state identities. Since we are interested in zero-temperature theory, we may set $\epsilon_4 \rightarrow \infty$. This leads to:
\begin{align}
    &I(\epsilon_1, \epsilon_2, \epsilon_3,\epsilon_4=\infty; a_1, \dots, a_4)\nonumber\\
    =& \exp -\left[a_1a_2 \mathrm{e}^{-\epsilon_1} +a_2a_3 \mathrm{e}^{-\epsilon_2}+ a_3a_4 \mathrm{e}^{-\epsilon_3} +a_1a_3 \mathrm{e}^{-\epsilon_1-\epsilon_2} + a_2a_4 \mathrm{e}^{-\epsilon_2-\epsilon_3}+a_1a_4 \mathrm{e}^{-\epsilon_1-\epsilon_2-\epsilon_3}+\sum_i a_i^2/2\right]
\end{align}
Using this result, we now compute the factor associated with the fourth order processes that involve a sequence of displacements of a bond phonon described by $\eta_1\eta_2\eta_3\eta_4$, where $\eta_i =\pm$ corresponding to an electron jumping in or out of the bond site. Note that $\sum_{i=1}^4 \eta_i = 0$ in order to go back to the ground state manifold. We then obtain
\begin{align}
    &G_{\eta_1\eta_2\eta_3\eta_4}(\xi_1,\xi_2) \nonumber\\
    \equiv &\sum_{\{n_i\}} \frac{\langle 0 | D(\eta_1\sqrt{X/2})|n_1\rangle \langle n_1| D(\eta_2\sqrt{X/2})|n_2\rangle \langle n_2| D(\eta_3\sqrt{X/2})|n_3\rangle \langle n_3|D(\eta_4\sqrt{X/2})|0\rangle }{(1+n_1/\xi_1)(1+n_2/\xi_2)(1+n_3/\xi_1)}\nonumber\\
    =& \xi_1^2\xi_2\int_0^\infty \mathrm{d}t_1\mathrm{d}t_2\mathrm{d}t_3  \exp\left\{-\xi_1 (t_1+t_3) - \xi_2 t_2 - \frac{X}{2}\left[\eta_1\eta_2 \left(e_1 + e_3\right) + \eta_2\eta_3  \left(e_2 + e_1e_2e_3\right) + \eta_1\eta_3  \left(e_1e_2 + e_2e_3\right) +2 \right]\right\}
\end{align}
where $e_i \equiv \mathrm{e}^{-t_i}$. In anti-adiabatic limit $\omega\rightarrow \infty  $, $X, \xi_i\rightarrow 0 $, $G\rightarrow 1$; In adiabatic limit $\omega\rightarrow 0  $, $X, \xi_i\rightarrow \infty $, $G\rightarrow \frac{1}{[1+X/(2\xi_1)]^2}\frac{1}{1+(1+\eta_1\eta_2)X/\xi_2}$. 

For the virtual processes where the intermediate state only has only phonon excitation (but no electronic excitation), we will also need to evaluate:
\begin{align}
    &G'_{\eta_1,-\eta_1,\eta_3,-\eta_3}(\xi) \nonumber\\
    \equiv &\sum_{\{n_i\}} \frac{\langle 0 | D(\eta_1\sqrt{X/2})|n_1\rangle \langle n_1| D(\eta_2\sqrt{X/2})|n_2\rangle \langle n_2| D(\eta_3\sqrt{X/2})|n_3\rangle \langle n_3|D(\eta_4\sqrt{X/2})|0\rangle }{(1+n_1/\xi) n_2/\xi (1+n_3/\xi)}\nonumber\\
    =& \xi \left.\frac{\partial G_{\eta_1,-\eta_1,\eta_3,-\eta_3} (\xi,
    \xi_2) }{\partial \xi_2} \right|_{\xi_2\rightarrow 0 } 
\end{align}
In anti-adiabatic limit $\omega\rightarrow \infty  $, $X, \xi_i\rightarrow 0 $, $G\rightarrow \frac{X}{2}$; In adiabatic limit $\omega\rightarrow 0  $, $X, \xi_i\rightarrow \infty $, $G\rightarrow 0 $ (no slower than $1/X^2$). 

In the main text, we call $G_+\equiv G_{++--} = G_{--++}$, and $G_- \equiv G_{+-+-} = G_{-+-+}$.

For the evaluation of the effective coeffecients for the dimer model, we need to compute
\begin{align}
    g_1 (\xi) \equiv &\sum_{\{n_i\}} \frac{\langle 0 | D(\sqrt{X/2})|n\rangle \langle n| D(-\sqrt{X/2})|0\rangle \langle 0| D(\sqrt{X/2})|n'\rangle \langle n'|D(-\sqrt{X/2})|0\rangle }{(1+n/\xi)^2  (1+n'/\xi)}\nonumber\\
    = & \xi^3\int_0^\infty \mathrm{d}t_1\mathrm{d}t_2\mathrm{d}t_3  \exp\left\{-\xi (t_1+t_2+t_3) - X + \frac{X}{2}\left[e_1e_2 + e_3\right]\right\} \\
    g_{2} (\xi_1,\xi_2) \equiv &\sum_{n,n'} \frac{\langle 0 | D(\sqrt{X/2})|n\rangle \langle n| D(-\sqrt{X/2})|0\rangle \langle 0| D(\sqrt{X/2})|n'\rangle \langle n'|D(-\sqrt{X/2})|0\rangle }{(1+n/\xi_1)^2  (1+n/\xi_2+n'/\xi_2)}\nonumber\\
    = & \xi_1^2\xi_2\int_0^\infty \mathrm{d}t_1\mathrm{d}t_2\mathrm{d}t_3  \exp\left\{-\xi_1 (t_1+t_3) - \xi_2 t_2 - X + \frac{X}{2}\left[e_1e_2e_3 + e_2\right]\right\} \\
    g_{3\pm} (\xi_1,\xi_2) \equiv &  \mathrm{e}^{(1\mp 1)X}\sum_{n,n'} \frac{\langle 0 | D(\sqrt{X/2})|n\rangle \langle n| D(\mp\sqrt{X/2})|0\rangle \langle 0| D(\sqrt{X/2})|n'\rangle \langle n'|D(\mp\sqrt{X/2})|0\rangle }{(1+n/\xi_1)  (1+n/\xi_2+n'/\xi_2)(1+n' /\xi_1)}\nonumber\\
    = & \xi_1^2\xi_2\int_0^\infty \mathrm{d}t_1\mathrm{d}t_2\mathrm{d}t_3  \exp\left\{-\xi_1 (t_1+t_3) - \xi_2 t_2 \mp X \pm \frac{X}{2}\left[e_1e_2 + e_2e_3\right]\right\} 
\end{align}
In anti-adiabatic limit $\omega\rightarrow \infty  $, $X, \xi\rightarrow 0 $, $g_i\rightarrow \mathrm{e}^{\mp X/2} \sim 1$; In adiabatic limit $\omega\rightarrow 0 $, $X, \xi\rightarrow \infty $, so $g_1\rightarrow \frac{1}{[1+ X/(2\xi)]^3}$, $g_2\rightarrow \frac{1}{[1+ X/(2\xi_1)]^2}\frac{1}{1+ X/\xi_2}$, and $g_{3\pm}\rightarrow \frac{1}{[1\pm X/(2\xi_1)]^2}\frac{1}{1\pm X/\xi_2}$. [For $g_{3-}$, the above adiabatic result only applies when $X<\xi_2, 2\xi_1$. For the cases studied in this paper, $\xi_2 =X $, $\xi_1 = W = \Delta_{12}/\omega_0 > X$, and $g_{3-} \rightarrow  \frac{\sqrt{\pi X/2}}{[1-X/(2W)]^2} $ in the adiabatic limit. ]

To keep the expressions short, we have made variable substitution $z_i \equiv \mathrm{e}^{-t_i}$ in Table.~\ref{effectivecoefficients}.

\section{The pseudo-spin $SU(2)$ symmetry}

In the anti-adiabatic limit $X\rightarrow 0$, the displacement operators $\hat{D}$ play no role in the transformed Hamiltonian $\hat{H}_T$, which will become simply a Hubbard model:
\begin{align}\label{transformedHamiltonian}
    \hat{U}^\dagger \hat{H} \hat{U}
    = & - t \sum_{\langle ij \rangle \sigma }  \left[ \hat{f}^\dagger_{\langle ij \rangle\sigma} \left(\hat{c}_{i\sigma}+\hat{c}_{j\sigma}\right)+ \text{h.c.}\right] + U_c \sum_{i} \hat{n}_{i\uparrow} \hat{n}_{i\downarrow} \nonumber \\
    & \ \ + \sum_{\langle ij \rangle } \left[\mathcal{E} \hat{n}_{\langle ij\rangle} - \frac{\Ueph}{2} \hat{n}_{\langle ij\rangle}^2\right]  +U_f \sum_{\lij } \hat{n}_{\lij \uparrow} \hat{n}_{\lij \downarrow}
\end{align}
where we have explicitly included other possible Hubbard interactions, $U_c$ and $U_f$, in order to generalize the results here.

Note that for arbitrary $t$, when $2\mathcal{E}+U_f = 2\Ueph + U_c$, this Hamiltonian has an  pseudo-spin $SU(2)$ symmetry generated by the generators:
\begin{align}
    \hat{J}_z &\equiv \sum_{i} \left(\hat{n}_i-1\right)+\sum_{\lij} \left(\hat{n}_{\lij}-1\right),\nonumber \\
    \hat{J}_- &\equiv \sum_{i} \hat{c}_{i\uparrow} \hat{c}_{i\downarrow} -\sum_{\lij} \hat{f}_{\lij \uparrow}\hat{f}_{\lij \downarrow} \ , \ \ \hat{J}_+ \equiv \hat{J}_-^\dagger 
\end{align}
These generators satisfy an $SU(2)$ algebra, and they are eigen-operators~\cite{PhysRevLett.63.2144} of the Hamiltonian in the sense that:
\begin{align}
    \left[\hat{H}_T, J_z\right] =0 \ , \ \ \left[\hat{H}_T, J_\pm\right] = \pm (-2\mu + U_c) J_\pm
\end{align}
where $\mu $ is the chemical potential that was left as implicit in the original expression of the Hamiltonian. Therefore, the Hamiltonian conserves both the $z$ component and the total pseudo-spin: 
\begin{align}
      \hat{J}_x \equiv (\hat{J}_+ + \hat{J}_-)/2 \ , \ \ \hat{J}_y \equiv (\hat{J}_+ - \hat{J}_-)/(2\mathrm{i})\ , \ \ \left[\hat{H}_T, \hat{J}_x^2+\hat{J}_y^2+\hat{J}_z^2\right] =0
\end{align}
This symmetry is a generalization of the pseudo-spin symmetry of the Hubbard model on bipartite lattices~\cite{PhysRevLett.65.120}. While we emphasize that the symmetry in the current model is present for arbitrary lattices (since $f$ and $c$ orbitals are always bipartite). 

When $2\mu = U_c$ (corresponding to ``half-filling'', one electron per orbital or $n = 1+z/2$), this symmetry become a full $SU(2)$ symmetry in the sense that all the three components of $\hat{J}$ are conserved, and it implies that the charge density operator $\hat{n}$ and local pairing operator $\hat{c}_{\uparrow} \hat{c}_\downarrow$ (or $-\hat{f}_\uparrow \hat{f}_\downarrow$) are symmetric. When away from ``half-filling'', only $J_z$ and $J^2$ are conserved. In this case, we note that the Hamiltonian takes the form $H_{SU(2)} + B_z J_z $ where $H_{SU(2)}$ is the fully $SU(2)$ symmetric part of the Hamiltonian and $B_z = (-2\mu +U_c )$ is the effective pseudo-Zeeman field. This implies that, all the energy levels can be written as $E_n(J,J_z) = E_n(J) + J_z B_z $. Therefore, for the ground state of the system, $|J_z| = J$ must be satisfied in order to fully optimize the pseudo-Zeeman energy for any $B_z$. This means that, in the ground state:
\begin{align}
    &J(J+1) = \vec{J}^2 = J_z^2+J_x^2+J_y^2 \implies J_x^2+J_y^2 = |J_z|
\end{align}
For the case we are considering, this becomes:
\begin{align}
    \left| \sum_{i} \hat{c}_{i\uparrow} \hat{c}_{i\downarrow} -\sum_{\lij} \hat{f}_{\lij \uparrow} \hat{f}_{\lij \downarrow} \right|^2  = (1+z/2 - n) N_{\text{site}}
\end{align}
where $n$ is the electron density per site, and $z$ is the coordination number. Dividing both sides by $N_{\text{site}}$ and take thermodynamic limit $N_{\text{site}}\rightarrow \infty$, we reach the conclusion:
\begin{align}
    \frac{1}{N^2_{\text{site}}}\left| \sum_{i} \hat{c}_{i\uparrow} \hat{c}_{i\downarrow} -\sum_{\lij} \hat{f}_{\lij \uparrow} \hat{f}_{\lij \downarrow} \right|^2 \rightarrow 0 
\end{align}
There are three possible physical conclusions from this result: 1. the ground state does not have off-diagonal long-range order (ODLRO); 2. it has uniform ODLRO, but $\langle 2 \hat{c}_{i\uparrow} \hat{c}_{i\downarrow} -\sum_j  \hat{f}_{\lij \uparrow} \hat{f}_{\lij \downarrow}\rangle=0$ is exact for any $i$; 3. it has ODLRO, but the order parameter oscillate in space and average to zero, i.e. it is a pair density wave (PDW) state~\cite{agterberg2020physics}. Therefore, this class of models may host PDW states, which remains to be explored in future studies. 

\section{The Hartree-Fock mean-field analysis on the weakly interacting model}

In the case of $\Ueph<\mathcal{E}$, the effective model is expressed in terms of site electrons ($c$-electrons). To fourth order in $t$, the effective Hamiltonian is
\begin{align}\label{cHamiltonian}
    \hat{H}_{c} = &- t_\text{eff} \sum_{\lij \sigma}\left(\hat{c}^\dagger_{i\sigma} \hat{c}_{j\sigma} + \text{h.c.} \right) - 2 J_\text{eff} \sum_{\lij }\hat{n}_{[i+j]\uparrow} \hat{n}_{[i+j]\downarrow}
\end{align}
where $\hat{n}_{[i+j]\sigma}
\equiv \left(\hat{c}^\dagger_{i\sigma}+ \hat{c}^\dagger_{j\sigma}\right)\left(\hat{c}_{i\sigma}+ \hat{c}_{j\sigma}\right)/2$ is the number of electrons in a bonding orbital between sites $i$ and $j$. Since any virtual movement of electrons necessarily costs $\mathcal{E}_1$ in the intermediate state, the expansion is valid as long as $|t|\ll \mathcal{E}_1$. [Note that, this condition does not require $|t|\ll \Ueph$.] 
As can be seen from Table \ref{effectivecoefficients}, $t_\text{eff}$ is second order in $t$ and $J_\text{eff}$ is fourth order.  Thus,  we should consider this theory in its weak coupling limit, $J_\text{eff}\ll t_\text{eff}$. This interaction term on each bond can be decomposed into a sum of an AF coupling, a pair hopping term, and a density attraction.

We analyze the weak-coupling instabilities in the context of a Hartree-Fock mean-field analysis that is reasonable in this limit.  We find that, on square and triangular lattices, $s$-wave SC is always the dominant instability for $n<2$. However, when $n\approx 1$ on the square lattice, there is also a $d$-wave pairing state that is meta-stable. To understand the reason for this, one may rewrite the interaction in the form $\frac{1}{N_{\text{site}}}\sum_{\bm{q},\bm{k},\bm{k'}}V_{\bm{q};\bm{k}\bm{k'}} \hat{c}^\dagger_{\bm{k}\downarrow}\hat{c}^\dagger_{\bm{q}-\bm{k}\uparrow}\hat{c}_{\bm{q}-\bm{k'}\uparrow} \hat{c}_{\bm{k'}\downarrow}$. For $\bm{q}=\bm{0}$ and for $\bm{k},\bm{k'}$ near the Fermi surface at $n\approx1$, we obtain:
\begin{align}
V_{\bm{k}\bm{k'}} &\approx -2J_{\text{eff}}\left[2+\cos k_x \cos k'_x + \cos k_y \cos k'_y\right] \nonumber\\
&= -J_{\text{eff}}\sum_af^{(a)}(\bm{k})f^{(a)}(\bm{k'}) 
\end{align}
where in the factorized form, the sum runs over an s-wave channel ($f^{(s)}(\bm{k})=2$), an extended s-wave channel ($f^{(\bar s)}(\bm{k})=\cos k_x+\cos k_y$), and a d-wave channel ($f^{(d)}(\bm{k})=\cos k_x-\cos k_y$). When $n\approx 1$, where the system has a large density of states near the two $X$ points, $(\pi,0)$ and $(0,\pi)$, of the Brillouin zone of the square lattice, the $d_{x^2-y^2}$-wave channel with form factor $f^{(d)}(\bm{k})$ is only moderately subdominant to the dominant $s$-wave channel  (which generically is a combination of $s$ and $\bar s$).  Moreover, the competition between the $s$- and $d$-wave paired states can be tuned by an additional weak on-site Hubbard repulsion, $U_c$; when $U_c \gtrsim 2.2 J_\text{eff}$, the $d$-wave paired state has the lower variational energy. Exactly at $n=1$, there is also an AF instability, which is also subdominant to the $s$-wave SC instability, but which is favored over all superconducting states when $U_c > 2 J_\text{eff}$.

\end{document}